\documentclass[12pt]{article}
\usepackage{graphicx}
\begin{document}
\begin{center}
{\Large\bf A study of obliquely propagating longitudinal\\
shear waves in a periodic laminate}\\
\vskip .25in
{\sl John Willis\\
DAMTP, Cambridge}
\end{center}
\vskip .25in
The basic purpose of this work is to demonstrate, by quite simple and explicit calculation,
the possibility that a simple laminate composed of ordinary materials can display the kind
of response associated with a ``metamaterial''. Specifically, a wave of Floquet--Bloch type
is analyzed, taking careful account of the requirements of causality and passivity. Then,
the generation of such a wave by transmission, from ordinary homogeneous material, into a
half-space composed of the laminate is considered. As is well-known for any dispersive
material, phase velocity is distinct from group velocity, and information is carried by
the latter. Thus, in line with the discussion of photonic crystals by Notomi [1],
refraction is considered
in terms of the group velocity. Negative refraction appears never to occur, if the
interface between ordinary material and composite is parallel to the interfaces in the
composite but negative refraction is possible, over an appreciable frequency range, if
the interface between ordinary material and composite is orthogonal to the interfaces
in the composite.  

\section{Basic ideas}
The methodology is as described in my document [2]. The material is elastic, and is periodically laminated, with period
$h$, in the $x_1$-direction. A ``transfer matrix'' method is used, which assumes a wave
which has only a $3$-component of displacement $u_3$, which has the form
\begin{equation}
u_3(x_1,x_2,t) = w(x_1)e^{(st + k_1x_1 + k_2x_2)},
\end{equation}
where $w(x_1)$ is periodic with period $h$. The reciprocal time parameter $s$ is chosen
in the form
\begin{equation}
s = \varepsilon - i\omega,
\end{equation}
where the radian frequency $\omega$ (corresponding to frequency $f$ so that
$\omega = 2\pi f$) will take a selected range of values while $\varepsilon$ is fixed,
small and positive. In the basic formulation, $k_2$ is chosen to have the form
\begin{equation}
k_2 = -s\,s_2,
\end{equation}
where the ``vertical'' component of slowness $s_2$ is fixed. The transfer matrix method
then delivers an equation that is easily solved for the Floquet--Bloch multiplier
$e^{k_1h}$ and hence $k_1$. If $k_1$ is a solution, then so is $\pm\{k_1 + 2n\pi i/h\}$
for any integer $n$. If $k_1$ has negative real part, the wave decays as $x_1$ increases.
It is reassuring to be able to verify that the associated group velocity
$\partial\omega/\partial k_1$ is positive in this case, consistent with the requirements
of causality and passivity. If the real part of $k_1$ is positive, the wave decays as
$x_1$ decreases and the group velocity is negative. A sample result is shown in Figure 1.
This is for a composite made from three isotropic materials with the following shear
moduli (GPa) and densities (kg\,m$^{-3}$):\\
$$\mu_1 = 200,\;\;\rho_1 = 8000,\;\;\mu_2 = 20,\;\;\rho_2 = 1100,\;\;\mu_3 = 8,\;\;\rho_3 = 1180.$$
It is convenient to regard ``material 3'' as being in the middle of a period cell surrounded by
equal thicknesses of ``material 2'', with this structural unit surrounded by equal thicknesses
of ``material 1'', as illustrated in Figure 2a. With this choice of periodic cell, the outer
layers of material 1 have thicknesses $0.2$ mm, the layers of material 2 have thicknesses $0.5$ mm
and material 3 has thickness $2.9$ mm, so that the period $h$ is $4.3$ mm. Figure 2b shows an
alternative (also symmetric) choice of unit cell for the same laminate.
Figure 1 plots the normalized
wavenumber variable $k_1h$ against frequency $f$, chosen in the conventional way so that ${\rm Im}(k_1h)$
lies between $0$ and $\pi$. The ``vertical'' slowness $s_2$ was chosen to be $s_2 = 9.0798 \times 10^{-5}$
s\,m$^{-1}$. The first stop band lies between approximately 240 and 400 kHz. The next commences at about
550 kHz. The branch of the second pass band, in between, is chosen so that $0\leq {\rm Im}(k_1h) <\pi$
so that the associated group velocity is negative.
\begin{figure}
\centering
\includegraphics[scale=0.4]{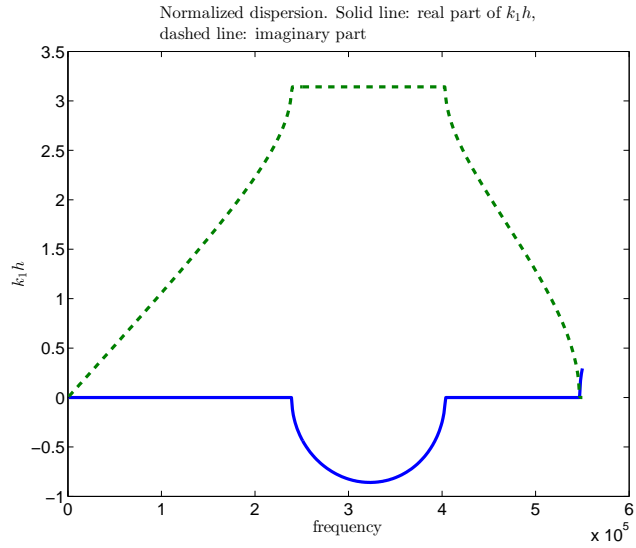}
\caption{\small First two branches of normalized dispersion relation at fixed $s_2$.
Solid line: real part, dashed line: imaginary part.}
\end{figure}
\begin{figure}
\centering
\includegraphics[scale=0.3]{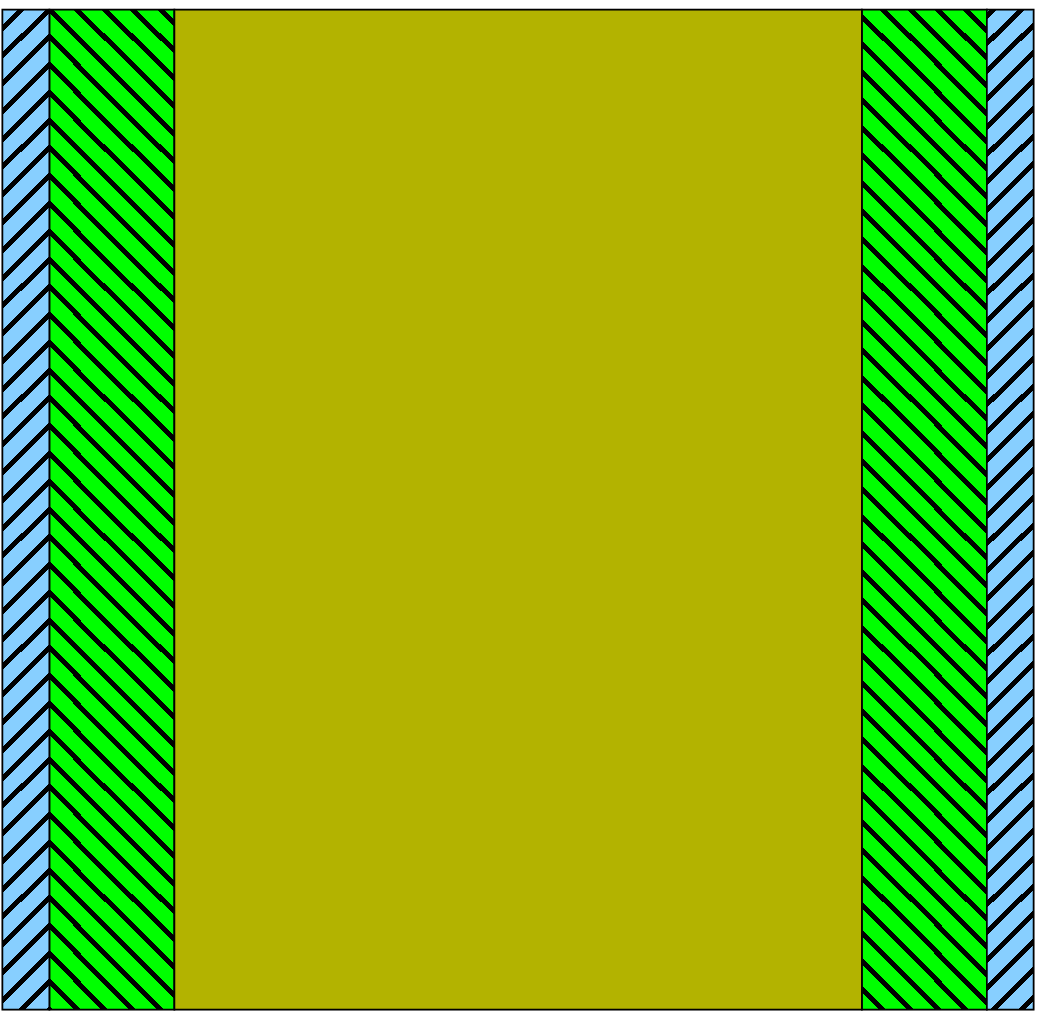} \hskip 1in
\includegraphics[scale=0.3]{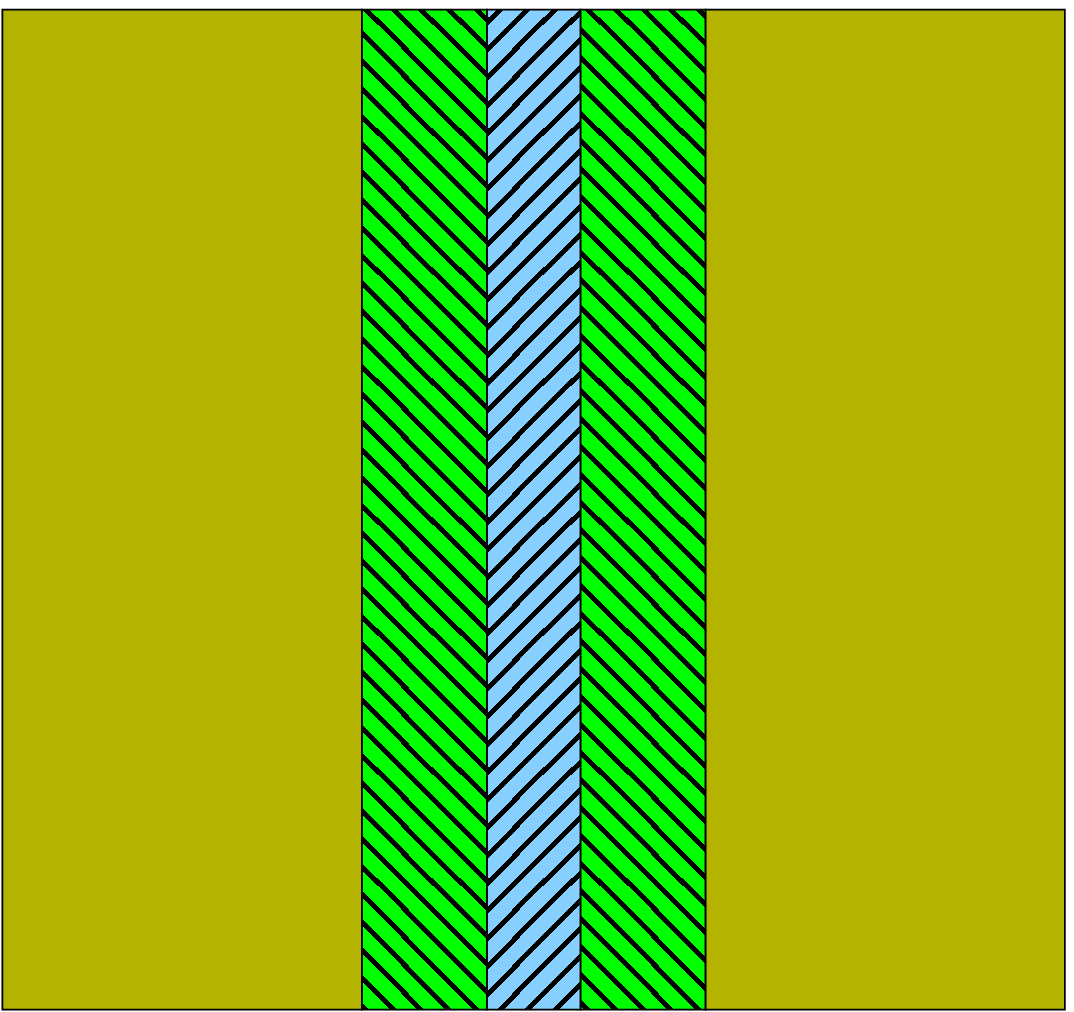}\\
(a)\hskip 2.1in (b)
\caption{\small Alternative choices for a basic period of the composite.}
\end{figure}
\section{Refraction}
Consider now what might happen if a plane wave were incident on a half-space of this laminated material.
First, suppose that the half-space occupies $x_1 >0$, and that the incident wave is carried by homogeneous
material, occupying $x_1 < 0$, at such an angle that the $2$-component of slowness is $s_2$ as specified
above. The wave transmitted into the composite will conform to the dispersion relation shown in Figure 1,
provided the frequency is less than 400 kHz. It will be a propagating (time-harmonic) wave up to frequency
240 kHz, and thereafter will be evanescent (with $k_1$ real and negative). For frequency between 400 and
550 kHz, the wave will be described by the branch of the dispersion relation that is {\it not} shown in
Figure 1: the imaginary part of the
wavenumber will either be negative or will be such that ${\rm Im}(k_1 h) > \pi$, so that energy will
be carried away from the interface $x_1= 0$. Figure 3 gives one acceptable choice for the
dispersion relation. The exact wavenumber and phase speed cannot be defined
uniquely, though the wave itself is unique. Figure 4 gives a plot of the group velocity associated with
Figure 3. The values in the stop band have no simple physical meaning and should be ignored. Note
that both components of group velocity are positive as would be expected, in both the first and second pass
bands.
\begin{figure}
\centering
\includegraphics[scale=0.4]{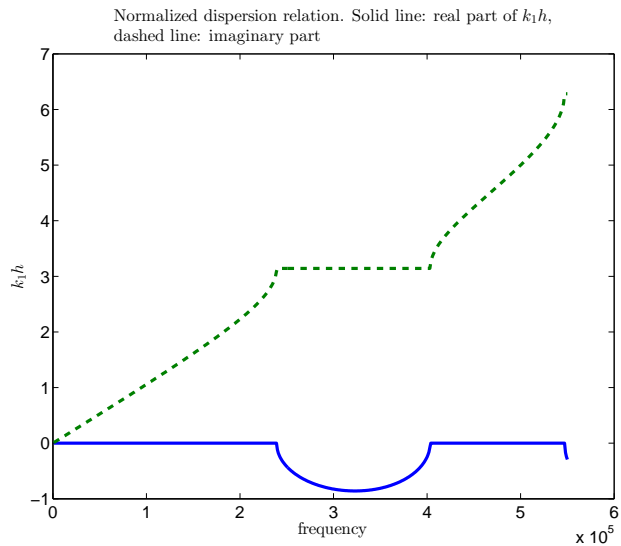}
\caption{\small First two branches of normalized dispersion relation at fixed $s_2$
 with positive 1-component of group velocity.
 Solid line: real part, dashed line: imaginary part.}
\end{figure}
\begin{figure}
\centering
\includegraphics[scale=0.4]{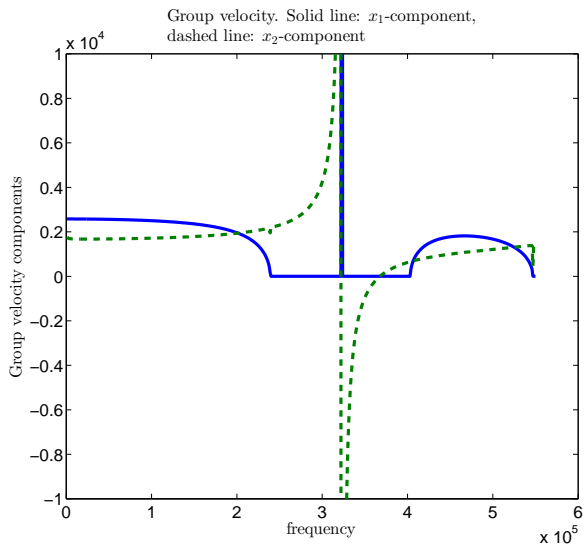}
\caption{\small Group velocity corresponding to Figure 3. Solid line: real part, dashed line:
imaginary part.}
\end{figure}

Now, however, consider what happens if the laminate occupies the half-space $x_2>0$ and a wave is incident
through uniform elastic material occupying $x_2 < 0$. It is appropriate this time to fix the 1-component
of slowness, $s_1$, and hence it becomes necessary to solve the dispersion relation for $k_2$, given
$\omega$ and $k_1 = i\omega s_1$. This is done in two stages, in effect, by inverting the relation
$k_1 = k_1(k_2,\omega)$. 
It is of particular interest to choose a value for $s_1$ that will generate
a solution in the second pass band shown in Figure 1. Results are shown next, for which
$s_1 = 2.1519$ s\,m$^{-1}$. This gives a solution point for $k_2$ that is close to that chosen for Figure 1,
at a frequency around 430 kHz. The resulting wave does not provide the {\it exact} description of the transmitted
disturbance because it cannot satisfy the interface conditions exactly. It is my intention to study
this further, partly along the lines developed for the purely one-dimensional problem in my document [3].
A plot of $k_2h$ versus frequency is shown in Figure 5. For
the frequency range that is plotted, $0.7421\pi \leq \omega s_1 h \leq 0.9623\pi$. The real part
(solid line) is negative for frequencies below about 415 kHz, corresponding to a non-propagating
disturbance. Throughout the propagating
range that is shown, the $x_1$-component of the group velocity is negative, while the $x_2$-component is
positive. The transmitted wave thus suffers {\it negative refraction}. This is illustrated in Figure 6,
which shows the angle $\alpha$ made between the group velocity and the $x_2$-direction. The Figure shows two
curves. One is the angle as calculated from the group velocity. The other is the corresponding angle,
calculated directly from the mean energy flux (averaged over time and over the spatial period). The
calculations were done only to an accuracy sufficient to produce the pictures, and the agreement between
the angles calculated by both methods is both interesting and reassuring.
\begin{figure}
\centering
\includegraphics[scale=0.4]{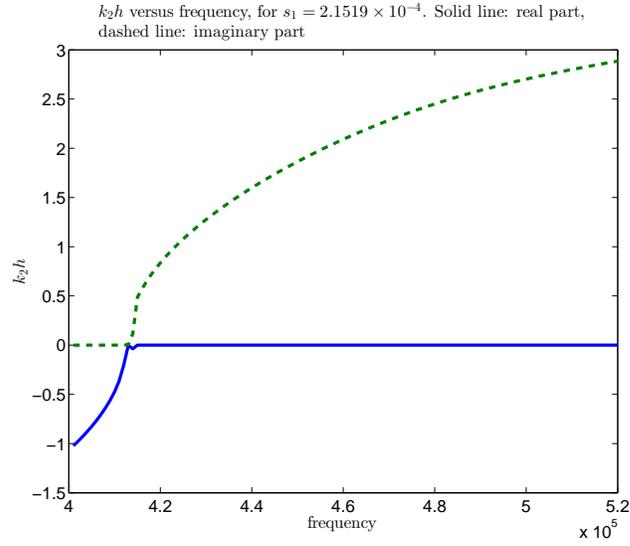}
\caption{\small First two branches of normalized dispersion relation at fixed $s_1$.
Solid line: real part, dashed line: imaginary part.}
\end{figure}
\begin{figure}
\centering
\includegraphics[scale=0.4]{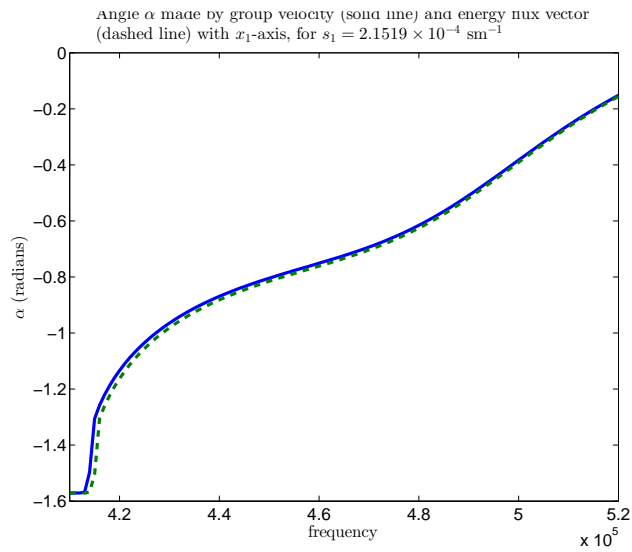}
\caption{\small Angle $\alpha$ made with $x_2$-axis. Solid line: group velocity, dashed line: energy flux.}
\end{figure}

The conclusions that can be drawn from this work are as follows. First, as is obvious, the laminate
is anisotropic  and responds anisotropically. Second, and perhaps quite significant, the calculations
demonstrate that a slab of material that is laminated in a transverse direction is capable of providing
negative refraction, over a significant frequency range. I have not attempted (yet) to make any
investigation of whether this material is ``homogenizable'', i.e. can be described in terms of
frequency-dependent tensors of moduli and density. 

\section*{References}
1. Notomi M. Theory of light propagation in strongly modified photonic crystals: Refractionlike
behavior in the vicinity of the photonic band gap. {\it Phys. Rev.} B 62, 10696--10705 (2000).\\
2. Willis J.R. Green's function for longitudinal shear in a periodic laminate. Privately circulated.\\
3. Willis J.R. Some thoughts on dynamic effective properties -- a working document. Privately circulated.\\

\section*{Addendum}
This report concentrated on 3-material laminates because that is where the investigation started.
However, I noted afterwards that two of the three materials that were employed in the example
are quite similar, so I re-did the calculation taking, ``material 2'' to have the same properties as
``material 3''. The resulting two-material composite can be described as having alternating layers,
as follows:

material 1: $\mu_1 = 200$\, GPa, $\rho_1 = 8000$\,kg\,m$^{-3}$, $h_1 = 0.4$\,mm,\\
material 2: $\mu_2 = 8$\,GPa, $\rho_2 = 1180$\,kg\,m$^{-3}$, $h_2 = 3.9$\,mm.

Figure 7 shows the resulting plot for the angle $\alpha$, which may be compared with Figure 7. The
feature that seems to be necessary is that layers of ``material 2'' are separated by thin,
relatively rigid and dense layers of ``material 1'' so that the total masses of each material are comparable.
\begin{figure}
\centering
\includegraphics[scale=0.4]{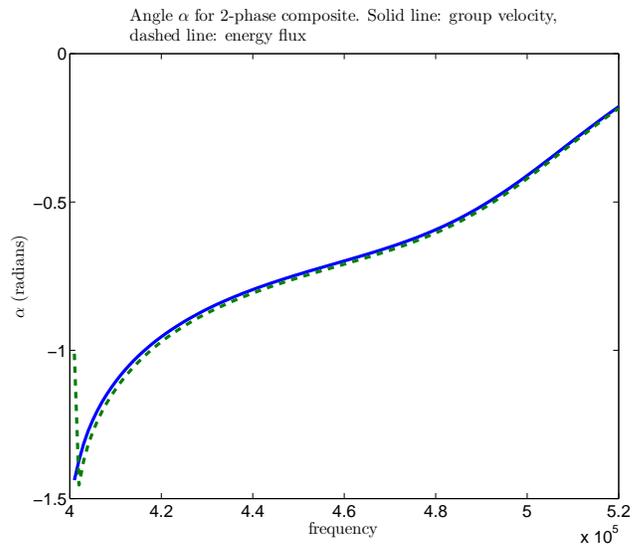}
\caption{\small Angle $\alpha$ made with $x_2$-axis, for the two-phase example.
Solid line: group velocity, dashed line: energy flux.}
\end{figure}

\end{document}